# 修例风波对有政治联系的香港上市公司的影响

学生姓名：＿＿＿王子奇＿＿＿

学　　号：＿2016312446＿

学　　院：中国金融发展研究院

专　　业：＿＿国际金融＿＿

指导教师：＿＿何重达＿＿

论文成绩：＿＿＿良＿＿＿＿

# 内 容 摘 要

香港修例风波对股市产生了重大影响，引起上市公司股价的较大变动。本文以每日游行人数作为抗议剧烈程度的衡量变量，发现在 2019 年 6 月 6 日至 2020 年 1 月 17 日间[①]，与泛民派相关联的公司股价受到抗议活动的额外负面影响。本文进一步发现，在《反蒙面法》实施之后，抗议活动对红筹股产生了额外的正面影响，对与泛民派相关联的公司产生了额外的负面影响。因此，本文认为在中央和香港特区政府采取严厉手段止暴制乱之后，红筹股的政治联系呈现了正面价值，而与泛民派的联系呈现出负面价值。

**关键词**：政治联系　股票市场　香港修例风波　政治风险

# ABSTRACT

Hong Kong's "anti-ELAB movement" had a significant impact on the stock market the stock price of listed companies. Using the number of protestors as the measurement of daily protesting intensity from 2019/6/6 to 2020/1/17, this paper documents that the stock price of listed companies associated with the pan-democratic parties were more negatively affected by protesting than other companies. Furthermore, this paper finds that after the implementation of the "anti-mask law", protesting had a positive impact on red chips but a negative impact on companies related to pan-democracy parties. Therefore, this paper believes that after the central government and the HKSAR government adopted strict measures to stop violence and chaos, the value of the political connection of red chips became positive while the value of the connection with pan-democracy parties became negative.

---

[①] 时间段的选取考虑到了数据的连续性、可用性，以及 2019-nCoV 的外生冲击，详见第二部分。

# 目　录

# 修例风波对有政治联系的香港上市公司的影响

2019 年 2 月，在中共中央纪委的倡导下，香港特区政府推进了《逃犯条例》修订草案，旨在填补司法漏洞，避免香港成为"逃犯天堂"。2019 年 3 月，香港"民间人权阵线"②开始针对《逃犯条例》组织大规模抗议游行，被称为修例风波。2019 年 6 月 16 日，游行人数达到了整个修例风波的最高峰，据"民间人权阵线"估计，200 万人参加了游行示威活动；据警方估计，33.8 万人参加了香港的游行活动。直至 2020 年 3 月，抗议活动还在继续。此次抗议活动使得暴力活动充斥着香港，对香港经济金融产生了及其恶劣的影响。香港股票市场作为香港金融业的重要组成部分，也受到了相当大的冲击。因此，本文旨在研究修例风波对香港股市的影响，并试图回答一个问题：有不同政治联系的香港上市公司会如何受到政治风险的影响？

本文隶属于政治不稳定性对金融的影响的研究领域。至少早在 Bernanke（1983），学者们就已经开始探究政治的不稳定是否会对金融领域产生实际上的冲击。股票市场作为金融领域的一个重要组成部分，也吸引了很多学者的兴趣。其中在理论分析层面，Pastor 和 Veronesi（2013）建立了股票价格对政治新闻反应的一般均衡政府政策选择模型。该模型指出，在较弱的经济条件下，政治不确定性带来的风险溢价幅度更大；政治不确定性降低了政府向市场提供的隐性看跌期权保护的价值。在实证研究中，也有学者发现了政治不稳定性对股票市场造成的冲击。在股票指数层面上，Ahmed (2015)发现，突尼斯"茉莉花革命"对突尼斯证券交易所的建筑业、工业、消费服务业、金融服务业、金融公司的行业指数和突尼斯指数的波动性的冲击是长期的。在公司层面上， David 和 Tiago（2019）发现，在 2017 年巴西总统丑闻中，与巴西国有开发银行有关联的公司，以及通过美国存托凭证交叉上市的公司的股价受此负面影响最大。

本文主要关注由于社会动荡所导致的政治不稳定性，例如抗议、游行、集会等。在这个研究方向上，Chau 等 （2014)发现，在 "阿拉伯之春"中，中东北非国家的伊斯兰指数的波动性显著增加，而抗议活动对传统市场的波动性几乎没有显著影响。Acemoglu（2018）发现，在埃及"阿拉伯之春"事件中，抗议越强烈（即抗议者人数越多），抗议所针对的掌权集团相关的公司股票价值越低于非关联公司的股票价值。本文受到Acemoglu（2018）的启发，使用每日抗议人数作为衡量香港修例风波剧烈程度的变量。

---

② "民间人权阵线"，简称"民阵"，是香港一个专门组织游行集会活动的平台，几乎所有香港泛民派成员都参与其中。截止 2019 年 6 月，参与民间人权阵线的民间团体及政治团体数目达 48 个。民间人权阵线最为人熟悉的是于 2003 年起每年主办香港七一游行。。

为了研究香港修例风波的抗议运动对不同属性的港股公司的影响，本文参考了研究政治关联与公司金融的相关文献。在这一研究领域中，很多相关文献使用了股票市场回报率作为衡量政治关联企业价值的指标（Roberts，1990；Johnson and Mitton，2003）。Roberts（1990）研究了美国参议员 Henry Jackson 之死对与 Henry Jackson 有联系的公司和与其继任者 Sam Nunn 有联系的公司股票回报率产生了不同的影响，指出了政治联系反应在股票市场上的价值。Akey（2015）发现，在美国大选后，捐赠给获胜候选人的公司的股票异常股本回报率要高出 3%。

但研究香港上市公司政治联系的相关文献较少。其中，Chan 和 Wei（1996）使用了GARCH 模型发现，在香港回归之前，当有着不利于中英合作的政治消息出现时，恒生指数会受到负面的影响；而出现有利于中英合作的政治消息时，恒生指数会受到正面的影响。而香港红筹指数③不会显著地受到政治消息的影响，可以作为政治风险的避风港。Bliss 等（2018）发现，有政治关联的港股公司较无政治关联的公司拥有更低的债务成本。该文对政治关联的判断方法为：如果公司至少一名董事属于 2006 届的香港选举委员会（共 800 人），则视为有政治关联。因此，本文以两种方式划分港股政治关联：①是否有董监高成员属于香港选举会（2016 届），香港立法会（2016 届），香港区议会（2019 届）④，并据此将相关政治联系标记为是否属于建制派或泛民派；②是否与内地有联系，即公司属于 H 股，红筹股，内地政府控股或中资民营股。本文针对不同标签的公司，分析抗议活动对其股票回报率的影响。

根据相关文献，本文提出如下假设：

假设 1：由于抗议活动无法动摇"一国两制"的基本政治制度，抗议活动使得市场预期政府会采取对泛民派政党的压制，因此相比于其他公司，抗议活动会降低泛民派关联的公司股价。

假设 2：抗议活动使得市场预期中央和香港特区政府会采取各种手段止暴制乱，因此与内地相关联的公司和支持特区政府的建制派关联的公司得到额外的正面反馈。

为了研究修例风波对香港不同属性的上市公司的影响，本文主要使用了两种研究方法：事件研究法和面板回归模型。在主要研究之前，先通过事件研究法，试图研究各个重要政治事件中，各类政治联系体现在股票累计异常收益率（CAR）的作用。本文发现，在《反蒙面法》的出台期间，与泛民派相关联的公司的 CAR 显著低于其他公司，而红筹股和内地政府控股公司的 CAR 显著高于其他公司。在全时间段（2019 年 6 月 6 日至 2020

---

③ 红筹股指的是与中国内地相关的公司，或内地控股或主营业务在内地的公司。

④ 均为修例风波前最新一届。

年 1 月 17 日）中，与建制派关联的公司，红筹股以及内地政府控股公司股票异常收益率显著高于其他公司。根据事件研究法的结果和相关文献（Chan 和 Wei，1996；Acemoglu，2018），本文得出初步的结论：当有止暴制乱的政府政策出现时，与建制派关联的公司、红筹股与内地政府相关联的公司抗性较好，与泛民派相关的公司抗性较差。

本文进一步通过面板回归模型检验每日抗议活动剧烈程度对各类港股的额外影响。本文以每日抗议人数作为抗议活动剧烈程度的衡量标准，主要针对反对《逃犯条例》的抗议集会活动。研究发现，政治联系并不仅体现在短期的政治事件中影响了股票的收益率，而且在整个修例风波中也展示了作用。在全时间段内，抗议的剧烈程度对于港股有着普遍的负面冲击。泛民派关联的公司和中资民营股受到了额外的负面作用。相比而言，与建制派关联的公司、红筹股与内地政府关联的公司抗性较好。

此外，本文视《反蒙面法》作为划分时间段的标志性事件。在《反蒙面法》实施之前，抗议活动对港股的影响是负面的。有内地联系的股票受到了额外的负面冲击，体现出了政治的不确定性。而经《反蒙面法》之后，由于对政府止暴制乱的期待逐渐提升，市场视剧烈的抗议活动为政府进一步止暴制乱、恢复经济的催化剂，因此港股异常收益率受到抗议活动的影响由负转正。然而，与泛民派相关联的公司受到了抗议活动的负面影响。结合事件分析法的结果，可以得出结论，在政府明确止暴制乱的决心后，引发修例风波的政党产生的政治联系呈现出了负面价值。与此相反，经《反蒙面法》之后，抗议活动对红筹股产生了额外的正面效应，体现出市场对红筹股的情绪回调，以及判断逻辑的根本变化。即随着政府明确止暴制乱的决心后，剧烈抗议活动只会使得政府采取更严厉的措施来整治暴乱，此时与内地政治联系的价值便呈现出来，使得红筹股受到市场额外的正面反馈。此外，H 股，内地政府控股、中资民营股以及与建制派相关的公司没有显著受到抗议活动的额外反馈。

因此，本文得出结论，在修例风波初期，由于抗议活动无法动摇"一国两制"的根本政治制度，与引导抗议活动的泛民派的政治联系没有产生正面价值。由于市场的不确定性，而与内地相关的公司受到了抗议活动的负面冲击。随着《反蒙面法》的制定，政府止暴制乱的措施逐渐增加，红筹股的政治联系展现了正面价值，而与引导抗议活动的泛民派的政治联系产生了负面价值。

此外，本文通过占中事件，构建了一个针对港股对抗议活动的敏感程度的变量。并以此作为控制变量，进行稳健性检验。本文发现，在占中事件中受到抗议活动负面冲击较大的公司，在修例风波中反而受到的负面冲击较小。本文理解为公司通过吸取占中事件的经验，避免了再次受到类似的冲击。通过纳入这个变量作为控制变量，本文的结论

没有发生改变，即结果相对稳健。

本文的贡献在于：①本文首次研究了香港修例风波对香港股市产生的影响。②本文将研究政治风险对金融影响的文献进行了扩展，展现了与其他文献不同的环境和视角。③本文首次展示香港政党对金融和股市的影响。④本文分析了在政治风险中，不同政治联系会展现出什么样的价值。

接下来本文的研究顺序如下：第一部分介绍修例风波和香港的政治、政党。第二部分介绍数据的收集和处理方法。第三部分介绍回归模型和结果。第四部分进行了稳健性检验。第五部分总结结论。

# 一、 修例风波与香港政治

## （一）"反对《逃犯条例》修订草案运动"

2018 年 2 月，香港男子陈同佳在台湾杀死女友潘晓颖后逃回香港，香港警方却不能以谋杀罪行进行起诉，引发群众热议[⑤]。该命案暴露出香港在逃犯引渡方面存在着司法漏洞。2019 年 2 月，在中共中央纪委的倡议下，香港特别行政区政府针对这次命案，提出推进《逃犯条例》修订草案。表示为了避免香港成为"逃犯天堂"，修订条例并填补司法漏洞。2019 年 3 月 26 日，林郑月娥会同行政会议，向立法会提交了《逃犯条例》草案[⑥]。然而其后引发一系列大规模抗议活动，即"反对《逃犯条例》修订草案运动"，简称修例风波或"反修例运动"。

从 2019 年 3 月开始 2020 年 3 月，修例风波一直在持续进行。大规模游行（超过 10 万人）主要由"民间人权阵线"引导。"民间人权阵线"总计组织了九次大规模游行。其中在 2019 年 6 月 16 日游行数量达到了整个修例风波的最高峰，主办方声称有 200 万人参加游行，警方声称有 33.8 万人参加游行。将警方与主办方的统计的数据取平均值，发现"民间人权阵线"引导的九次抗议活动的平均值达 50 万人，远超过所有抗议活动的平均值（约 4 万人）。"民间人权阵线"（简称民阵）是香港泛民派用于组织民间游行的平台，于 2002 年 9 月 13 日成立，以组织每年的香港"七一游行"为主要活动。民阵主要由泛民派政党及其所支持的民间团体构成，包括香港民主党、香港公民党、香港工党、香港职工会联盟、新民主同盟等。在修例风波中，泛民派的身影愈发明显，因此相关政党是本文的一个重要研究方向。

---

[⑤] 详见 https://baike.baidu.com/item/%E9%99%88%E5%90%8C%E4%BD%B3/23665287?fr=aladdin 。

[⑥] 全称《2019 年逃犯及刑事事宜相互法律协助法例（修订）条例草案》。

由于修例风波，香港特别行政区政府做出了相应的应对措施。2019 年 6 月 15 日，特首林郑月娥宣布，特区政府决定暂缓修订《逃犯条例》的工作。2019 年 10 月 23 日，香港立法会召开会议，保安局局长李家超宣布，政府决定撤回《逃犯条例》。为有利于止暴制乱，10 月 4 日，香港特区政府特别行政会议引用《紧急情况规例条例》（即紧急法），制定了《禁止蒙面规例》（即反蒙面法），并于 10 月 5 日开始实施。11 月 18 日，香港特别行政区高等法院裁定《紧急情况规例条例》部分条款不符合基本法，致使有关条款无效，使得《反蒙面法》失效。⑦

中共中央高度关注香港局势，并从做出了充分的研判和部署。2019 年 8 月 7 日，国务院港澳办和驻港联络办在深圳举行了香港局势座谈会。港澳办主任张晓明宣布了中共中央关于稳定香港局势的重要精神，指出修理风波已经向颜色革命方向发展。中央有足够的办法和力量，来平息可能出现的各种动乱，绝不容忍挑战"一国两制"原则底线的行为，绝不向反对派妥协退让。11 月 4 日，国家主席习近平在上海会见香港特首林郑月娥。习近平表示，林郑月娥带领特区政府恪尽职守，做了大量艰辛的工作，对其表示信任支持。希望香港各界人士贯彻"一国两制"方针和基本法，共同维护香港的繁荣稳定。11 月 19 日，依照基本法的有关规定，国务院任命邓炳强为新一任警务处处长，免去卢伟聪的警务处处长职务。12 月 6 日，公安部部长赵克志会见香港警务处处长邓炳强，强调了中央政府和公安部永远是香港警队的坚强后盾。

由于修例风波，香港经济于 2019 年下半年受到明显的负面影响。包括经济成长遭削减、股市受冲击、流动性收紧、营商环境恶化，企业活跃度降至环球金融危机以来的最低水平。香港 2019 年 GDP 下跌 1.2%，为自 2009 年以来首次。由于修例风波，惠普国际 24 年以来首次下调香港评级，并将展望评级调至"负面"。穆迪其后也将香港的评级展望调至"负面"，将评级调低至 Aa3。大规模罢工运动导致股市急跌，8 月 5 日首次"三罢"时恒生指数跌 2.85%，即 767 点；11 月 11 日第二次"三罢"时恒生指数跌 2.62%，即 724 点。

---

⑦ 2020 年 4 月 9 日，香港高等法院上诉法庭 9 日颁下判案书裁定，《紧急情况规例条例》("紧急法")中有，关行政长官会同行政会议认为在出现危害公安的情况时订立紧急情况规例的部分符合基本法规定。

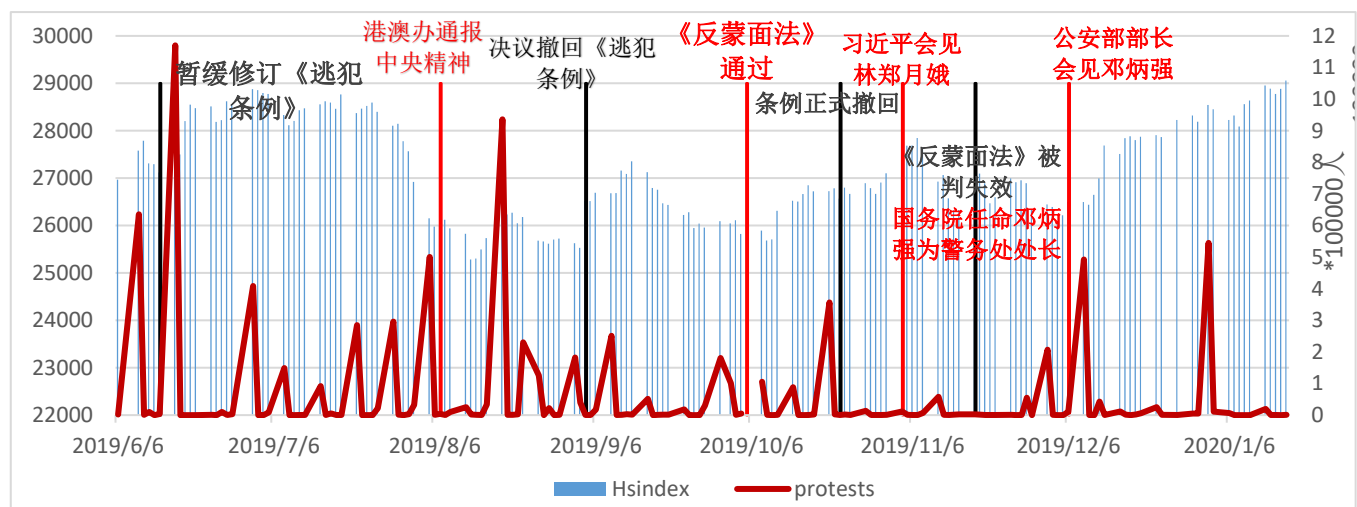

图 1：恒生指数、抗议人数及相关事件

### （二）香港政党与政治

政党是香港政治中一个不可忽视的角色，但并不是学者们研究的重点（殷俊&蔡幗仪，2019）。对于香港是否存在着政党政治，学者们持有不同态度。政党政治指"由政党掌权，而且政党在国家和社会生活中处于中心地位的政治结构"（王韶兴，2000）。而自 1997 年 7 月 1 日回归后，基本法的制度设计缩减了政党活动的空间（曹旭东，2013）。主要体现在：行政长官由选举委员会产生，而且选举委员会设有功能组别选举委员（不需要竞选），主要由建制派构成。此外，香港《行政长官选举条例》第 31 条规定：在选举中胜出的候选人不能是任何政党成员，并且在担任行政长官的任期内不参加任何香港政党。这就使得香港的政党无法掌握行政权，也就难以形成的一般意义上的政党政治。（殷俊&蔡幗仪，2019）虽然无法出现执政党，但是政党仍然会影响到香港的政治生活。即使单个政党的影响不足，但是政党形成的政治联盟却能影响到政治走向。因此，部分学者将其称为"半政党政治"（殷俊&蔡幗仪，2019；曹旭东，2013；黎沛文，2015）。

香港第一个政党"香港民主同盟"于 1990 年 4 月 23 日成立，同时也是第一个泛民派政党。90 年代，英国认为加速推动民化进程对于笼络香港民心，保留英国影响及维持国际舆论有重要作用，任命彭定康出任港督。"彭改计划"使得香港泛民派极大的发展，泛民派势力主要集中于立法局。在 1995 年至 1997 年，香港立法局由"政策质询型"走向"政策影响型"的议会（蔡子强，1996）。随着泛民派的势力大增，各个泛民派政党相继成立：民主党（1994 年，起源于香港民主同盟），公民党（2006 年），社会民主连线（2006 年），新民主同盟（2010 年），人民力量（2011 年），其口号多为"双普选"、

反对"二十三条"。这些泛民派政党形成了泛民派政治团体，在选举会、立法会和区议会中成为了一股政治力量（例如 2017 年选举会中的民主 300+）。与此同时，建制派政党也逐渐发展起来：自由党（1992 年，主要由工商界构成）、民主建港协进联盟（1992 年，简称民建盟）、新民党（2010 年）、香港经济民生联盟（2012 年，简称经民联）。建制派坚持"一国两制"并拥护基本法，与泛民派相对立。建制派和泛民派在香港特首选举委员会、香港立法会和香港区议会中占据重要地位。因此，本文将以这些参加香港政治（即三会）的政党党员所在的公司作为研究对象。

香港泛民派政党和政治团体曾多次引导香港民众集会游行。自 2003 年起，由泛民派组织成的"民间人权阵线"每年七月一日都会针对基本法 23 条组织大规模游行（香港七一游行）。2014 年，由泛民派和"学联"等社会组织引导了针对中央关于 2017 年特首选举方案的大游行，从 2014 年 9 月 26 日持续到了 2014 年 12 月 15 日结束，被称为"雨伞运动"或者"占中事件"。据主办方估计，9 月 28 日至 10 月 1 日每天游行人数多达 20 万人，整场运动参与人数达 120 万人。该事件对香港经济、交通和金融产生了巨大的影响，9 月 29 日当天（大规模游行后的第一个交易日），恒生指数下跌 1.89%。"占中事件"进一步催生了泛民派思潮，为之后的修理风波埋下了祸根。

## 二、 数据

### （一）数据来源与样本选取

本文涉及到的上街抗议数据来源于维基百科的统计，政党数据来源于香港特别行政区选举管理委员会官网和维基百科。其他数据，如股价数据，董监高人物数据，控制变量数据均来源于 Wind 数据库。

本文选取抗议时间段为 2019 年 6 月 6 日至 2020 年 1 月 17 日，因为考虑了数据可用性和 2019-nCoV 的外生冲击。总计 155 个交易日。为了因变量的计算和其他变量的获取，本文仅考虑在 2018 年之前上市且未在 2020 年 1 月 17 日前退市的港股，并经过去极值之后，共 1961 只港股纳入样本范围。

### （二）抗议剧烈程度

本文参考了 Acemoglu（2018），使用上街抗议人数作为衡量香港修例风波剧烈程度的变量。修例风波中的上街游行分为两种，一类是反对《逃犯条例》的游行，另一类是支持特区政府和警察的游行。本文主要考察第一类游行所造成的影响。数据来源于维基百科上的统计数据，并对这些数据进行了处理。考虑到数据的连续性和可得性，我们从 2019 年 6 月 6 日开始统计上街游行数据（见维基百科），并作为研究时间的开始。并以

2020 年 1 月 17 日结束，主要考虑到 2019-nCoV 的外生冲击。

首先，我们对估计型数据做出了类似于 Acemoglu（2018）的处理。出现"数以百计"、"数千"等表示方式，我们使用 500、5000 作为上街人数估计值。如果出现"过百"、"上千"等字样，我们采用 100、1000 作为估计值。针对大规模游行（10 万以上），以及出现警方和主办方数据估计不符时，我们采用各个渠道的估计平均值作为上街抗议人数。经处理后，本文将所有非交易日的抗议数据合并到下一个交易日，得到变量 protests。从 2019 年 6 月 6 日到 2020 年 1 月 17 日，香港证券交易所总计 155 个交易日。在 155 个交易日中，平均每个交易日的抗议人数达五万人。本文进一步处理数据，将 155 天的游行数据标准化，得到变量 stdprotests。变量名见表 1，统计结果详见表 2，变量相关性见图 2。

### （三）港股政治联系

1. **政党联系**

参照 Bliss 等（2018），并考虑到数据可用性（无法获取政党所有成员名单），本文把参加最新一届的选举会、立法会和区议会的成员考虑在内。2016 届选举会中，建制派占 805 人，泛民派占 325 人，总共选举会人数 1200 人，并最终选举林郑月娥成为香港特首⑧。2016 届立法会（LegGo）中，43 人为建制派人士，23 人为泛民派人士，总共立法会议员为 70 人。2019 届区议会中，建制派议员占 86 人，泛民派议员 389 人，全部区议会议员为 479 人。去掉重复议员后，总计 1588 人。根据议所在政党类型，参照维基百科本文将政党议员分为建制派和泛民派。

本文将政党议员名单与香港上市公司董监高人员进行匹配，设立标准为董监高至少有一人为政党议员，则视为与该政党有联系。最终经过匹配，发现有 310 家上市公司与建制派政党有关，有 72 家公司与泛民派政党有关。变量命名见表 1，统计结果见表 2，变量相关性见图 2。

2. **港股的内地联系**

除了针对政党的政治联系外，本文还考虑到由于公司本身性质产生的政治关系，即是否与内地有联系。本文参照了 Chan 和 Wei（1996），考虑了红筹股产生的政治联系。红筹股（变量名称：red）是指在中国境外注册、在香港上市的中国内地实体控股的公司的股票（内地实体包括国有企业及由国内的省、市机关所控制的实体）。分类来源于 Wind

---

⑧ 由于自由党党魁一直持有反对派立场，因此本文将自由党划为泛民派政党。

数据库。

此外，港股还有其他的内地联系。例如，H 股（变量名称：H）指注册地在内地、上市地在香港的中国股票。中资民营股（变量名称：chinaasset）指主要资产来源中国内地、实际控制人为中国内地公民，或主营业务收入来自中国内地的股票。因这一部分中有部分既属于 H 股也属于中资民营股，所以本文使用了 Wind 的专门区分的类型：中资民营股（非 H 股）。本文还将受中央或内地地方政府控制的股票切上标签：centralcontrol，视为公司与内地政府的关联。变量命名见表 1，统计结果见表 2，变量相关性见图 2。

**（四）因变量：AR&CAR**

本文参照了 Acemoglu（2018）中的事件研究法，以异常收益（AR）和累计超常收益率（CAR）为因变量进行研究。计算方法如下：

1. 选定事件前估计窗口期：2018 年 1 月-2018 年 12 月，目的为避开修例风波产生的影响。设定市场指数为 MSCI 香港指数，分别对每股进行 OLS 回归，计算出每股贝塔 $\beta_i$：$R_{it} = \alpha_i + \beta_i R_{Mt} + \varepsilon_{it}$

其中，$R_{it}$ 指股票 i 在窗口期 t 时间的股票收益率，$R_{Mt}$ 指市场指数在 t 时间的收益率，$\varepsilon_{it}$ 为股票 i 的残差项。

2. 在修例风波时间段（本文以 2019 年 6 月 6 日至 2020 年 1 月 17 日为研究时间），计算异常收益（AR）：$AR_{it} = R_{it} - \beta_i R_{Mt}$

3. 针对研究事件时间段[a, b]，计算累计超常收益率（CAR）：$CAR_i[a, b] = \sum_{t=a}^{b} AR_{it}$

**（五）衡量股票对抗议活动的敏感性**

本文参照了 Acemoglu（2018），研究香港上市的股票对抗议活动的敏感性。在修例风波 5 年前，2014 年，泛民派和"学联"等社会组织引导了针对中央关于 2017 年特首选举方案的大游行，被称为"雨伞运动"或者"占中事件"。本文将这段时间的抗议程度对香港股票的影响，作为港股对类似抗议活动的敏感性度量。具体计算方法如下：

1. 设定估计窗口期为 2014 年 9 月 26 日至 2014 年 12 月 15 日。

2. 通过对每一支股票进行 OLS 回归，计算占中事件之前的个股 $\beta_i$，窗口期为 2014 年 1 月-2014 年 6 月，市场指数为 MSCI 香港指数，按照计算因变量的相同方法计算占中事件中的异常收益率 $AR_{it}$。

3. 对每一个股票进行使用 OLS 回归：$AR_{it} = \alpha_i + \beta_i^{unrest} protests + \epsilon_{it}$

protests 是指占中事件中每一个交易日的抗议人数（处理方式同修例风波中的抗议人数），每一支股票得到一个受到占中事件抗议剧烈程度的敏感度 $\beta_i^{unrest}$，并记为横截面数据 occupybeta。变量命名见表 1，数据统计见表 2。

**（六）其他控制变量**

除主要研究变量外，为了控制其他因素对因变量的影响，本文采用了一些控制变量。并去掉了一些极端值。

总资产（size）：为以港元计价的公司总资产的对数。

资产负债率（leverage）：总负债/总资产，去掉总负债超过总资产的公司。

市盈率的倒数（1/PE）：EPS/股价，是面板类型数据。

换手率（turnover）：日内总交易量/流通股总量，去掉大于 1 的数据，是面板类型数据。

MSCI 世界指数（worldchange）：为了控制世界股票市场的走势影响。

hkbeta：2018 年 1 月 1 日-2018 年 12 月 31 日各股 beta，市场指数为 MSCI 香港指数。

worldbeta：2018 年 1 月 1 日-2018 年 12 月 31 日各股 beta，市场指数为 MSCI 世界指数。

变量命名见表 1，数据统计见表 2，变量相关性见图 2，

# 三、 实证研究

**（一）事件研究法**

本文采用了标准事件研究法，选择了 8 个重要政治事件进行分析。本文同 Acemoglu（2018），对各个重要事件中股票的累计异常收益率 CAR 受到的影响进行分析，并分别采用事件前一天到后一天的三天累计异常收益率 CAR[-1,1]，以及事件前两天到后两天的五天累计异常收益率 CAR[-2,2]。选择的政治事件见表 3。

模型如下：

$$CAR_i[-1,1] = \alpha_i + \gamma N_i + \delta X_i + \theta Z_s + \varepsilon_i$$

其中，$CAR_i[-1,1]$ 是指事件前后一天的三天累计异常收益率，也可为 CAR[-2,2]。$N_i$ 是指研究的政治联系类型，有变量 proestablish，pandemo，H，red，centrolcontrol，chinaasset。$X_i$ 为控制变量，包含 worldbeta，size，leverage。$Z_s$ 是指控制的行业固定效应。$\varepsilon_i$ 为残差项。以下展示了对股票 CAR 有着显著影响的事件。

事件一：2019 年 10 月 5 日《反蒙面法》开始实施

《反蒙面法》意味着香港特区政府通过紧急法的法律形式，对暴乱进行处理。 根据 4，与泛民派相关的公司（pandemo）的累计异常收益率受到负面的影响。而有着与中国内地政治联系的公司（H 股，红筹股，政府控股和中资民营股）与无联系的公司相

比，股票累计异常收益率更高。这说明了《反蒙面法》对香港暴乱的质量作用体现在了股票市场上，即使得泛民派相关公司股票受到了负面预期的冲击，而与内地关联的公司得到了股票市场正面的反馈，结果详见表4。

事件二：2019年10月23日《逃犯条例》正式撤回

在通过《反蒙面法》之后，中央继续采取多项行动解决香港问题。《逃犯条例》的撤回，是中央和香港特区政府为了缓解修例风波，解决香港民生问题做出的决定。根据表5，由于抗议活动是由泛民派挑起，在撤回《逃犯条例》时，与泛民派相关的股票相较其他股票有着较高的累计异常收益率。而中资民营股受到了短暂的负面冲击，结果见表5。

事件三：2019年12月6日，公安部部长赵克志会见香港警务处处长邓炳强

2019年12月6日，在与香港警务处处长邓炳强的会谈中，公安部部长赵克志强调了中央政府和公安部永远是香港警队的坚强后盾。该事件体现了中央政府对港警的支持和打击香港暴乱分子的决心。该事件中，H股（H）、红筹股（red）和内地政府控股的股票（centralcontrol）等与内地相关联的股票均受到了股票市场的正面反馈，结果见表6。

根据这三次事件的检验，本文得出初步结论：当出现打击泛民派主导的抗议活动的政治消息时，泛民派相关的股票会受到负面冲击，而与内地关联的股票会受到市场的正面反馈；在《逃犯条例》撤回事件中，泛民派相关的股票收益率得到上调。

根据事件研究法结果，以及《反蒙面法》的标志性作用，本文将时间段分为两个部分，即《反蒙面法》实施前和《反蒙面法》实施后。

本文分别计算了《反蒙面法》实施前以及《反蒙面法》实施后的累计异常收益率，和全时间段的累计异常收益率，进行研究，结果见表7。在修例风波全时间段中，与建制派有联系的公司，红筹股公司，内地政府控股的公司有着较其他股票更好的累计异常收益率。在《反蒙面法》实施前，红筹股和内地政府控股的公司受到的市场反馈较好；在《反蒙面法》实施后，红筹股和与建制派关联的公司受到的市场反馈较好。

由此说明，在修例风波中，红筹股的表现尤为出色，抗性也较好，其他与内地关联的公司和与建制派关联的公司也有一定的抗性。而当打击香港暴乱的政治消息出现时，与泛民派关联的公司会受到负面影响。

## （二）面板回归模型

为了研究每日的抗议活动剧烈程度对各类有政治联系的香港股票的影响，验证假设，本文采用了固定效应模型进行实证研究。并把港股每日受到的影响用异常收益率（AR）来刻画。主回归模型如下：

$$AR_{it} = \alpha_{it} + \beta_1 N_i + \beta_2 P_t + \gamma(P_t \times N_i) + \varphi X_{it} + \phi \eta_s + \varepsilon_{it}$$

其中，$AR_{it}$ 是指每日的个股异常收益率，计算方法见上文。$\alpha_{it}$ 是常数项。$N_i$ 是指用来区分不同种类政治联系的 dummy 变量，包含 proestablish，pandemo，H，red，centrolcontrol，chinaasset。$P_t$ 是经过标准化后的每日抗议人数，即变量 stdprotests。$(P_t \times N_i)$ 为交叉项。$X_{it}$ 是控制变量，包含 worldchange，size，leverage，1/PE，turnover，$AR_{it-1}$，$AR_{it-2}$。$\eta_s$ 是指控制的行业固定效应。$\varepsilon_{it}$ 是残差项。

由于本文主要关注的系数为 $\gamma$，因此，此模型按照常规假设 Cov$(P_t \times N_i, \varepsilon_{it} | X_{it}, \eta_s)$ =0 具体而言，本文假设：没有遗漏在时间序列上变化且与异常收益率和抗议人数相关的变量；有政治联系公司的每日异常收益率与抗议强度之间不存在反向因果关系。

回归结果见表 8。（1）至（7）列均使用了相同的控制变量并控制了行业固定效应。（1）列将标准化后的抗议人数作为研究变量，发现抗议剧烈程度的增加，会显著地降低港股异常收益率，即视为抗议活动对全部港股有着普遍的负面冲击。（2）至（3）列主要研究拥有党派联系产生的政治联系的港股公司所受修理风波的影响。结果表明，在全时间段内，相比于其他公司，游行人数的增加会更加负面影响与泛民派政党有联系的港股公司。而与建制派有联系的公司在全时间段内没有受到额外的负面影响。

（4）至（7）列研究修例风波对除政党以外的其他政治联系的作用。（4）至（5）列研究标的为 H 股和红筹股。结果发现，抗议活动没有对红筹股和 H 股产生额外的影响。而在整个时间段内，H 股的异常收益率低于其他股票，红筹股的异常收益率超过其他股票。（6）至（7）列表明，抗议剧烈程度会对中资民营股（非 H 股，即不在内地注册的公司）产生额外的负面影响。

在全时间段内，受到抗议活动额外冲击的主要为与泛民派有关的股票和中资民营股。而抗议活动对其他与内地关联的股票没有显著的额外影响。此外，红筹股在修例风波全时间段内有比其他股票更高的异常收益率。

## （三）分时间段研究

经过前文的事件研究法，本文发现，《反蒙面法》可以作为一个划分时间段的标志。制定《反蒙面法》是香港特区政府使用紧急法，在法律的层面上止暴制乱的决定。《反

蒙面法》实施后，中央愈发关注香港形式，并采取了相应的措施止暴制乱，撑警撑特首。因此，本文将其作为标志性事件，将研究的全时间段化为《反蒙面法》实施前，2019年 6 月 6 日至 2019 年 10 月 4 日；和《反蒙面法》实施后，2019 年 10 月 5 日至 2020 年 1 月 17 日。

表 9 展示了在《反蒙面法》实施前抗议活动对各类有政治联系的股票的影响。（1）列表明，在《反蒙面法》实施前，抗议活动对港股的异常收益率有着显著的负面影响。（3）列表明，与党派有关的公司在《反蒙面法》实施前股票异常收益率没有显著地受到抗议活动的影响。而（4）至（7）列的结果显示出，在《反蒙面法》实施前，抗议活动对红筹股、内地政府控股的公司和中资民营股的负面影响较其他股票更大。

表 10 展示了在《反蒙面法》实施后抗议活动对各类有政治联系的股票的影响。（1）列表明，在《反蒙面法》实施后，抗议活动对港股的异常收益率有着显著的正面影响。（3）列发现，在《反蒙面法》实施后，抗议活动对于泛民派相关的股票较其他股票有显著的负面效应。（5）列指出，在《反蒙面法》实施后，抗议活动对于红筹股有着额外的正面效应。

**（四）结果分析**

结合事件分析法和面板回归模型，本文从以下的方式来理解回归结果：

首先，关于抗议的剧烈程度对港股的普遍影响。本文认为，在《反蒙面法》实施之前，由于政治风险的不确定性和香港经济的低迷状态，投资者产生了恐慌情绪，因此对香港股市呈现悲观态度。抗议活动越剧烈，投资者对未来经济和股市的负面情绪越高。而《反蒙面法》出台之后，投资者看到了中央和香港特别行政区政府止暴制乱的决心。因此，当抗议活动越剧烈，对政府进一步采取措施止暴制乱、回复经济的期待性越强，因此对港股股市产生了普遍的正面效应。

其次，关于抗议活动对与香港政党有关的股票的额外效应。本文认为，相比于其他的抗议事件，例如"阿拉伯之春"，修例风波的特点是不会像其他事件一样产生变革政府的作用，"一国两制"的根本制度并不会受到抗议活动的影响。因此，结合香港的"半政党政治"，本文认为，与泛民派相关的公司并不会受到正面反馈，因为这些公司的泛民派政治联系无法给它们提供为了更好的寻租空间。这一点与 Acemoglu（2018）的结果不同，原因在于修例风波无法产生政权更迭的作用，因此与泛民派相关的政治联系毫无价值。而且在香港特区政府采取手段止暴制乱时，这些政治联系反而对公司产生不利影响，使得股价被低估。而与支持"一国两制"的建制派相关公司则不会受到抗议活动产生的额外影响。

最后，关于抗议活动对有内地关联的股票的额外效应。本文认为，内地联系在修例风波中展现的价值是有分化的。在修例风波期间，红筹股和内地政府控股的公司累计异常收益率显著高于其他公司。在《反蒙面法》出台之前，由于抗议活动的影响，除 H 股外的有内地联系的股价都受到了额外低估，而且 H 股本身在修例风波时间段中的股票价值偏低。本文认为，这是由于修例风波初期，产生了巨大的政治不确定性，市场对与内地联系的股票产生了悲观情绪。而在政府明确表示采取严厉的手段止暴制乱时，投资者认为红筹股未来的寻租能力会更强，红筹股的政治联系的价值便脱颖而出。因此，抗议活动越剧烈，政府的措施越严厉，红筹股受到的正面反馈越大。而中资民营股及其他的没有受到显著的额外正面效应，以至于在全时间段范围内，抗议活动对中资民营股产生了额外的负面效应。

## 四、 稳健性检验

### （一）港股对抗议活动的敏感性

在 2019 年修例风波的 5 年前，2014 年香港也爆发了一场抗议游行——占中事件（又称"雨伞运动"）。本文参照 Acemoglu（2018）构造埃及上市股票的对抗议活动的正向敏感性$\beta^{unrest}$，构造了 occupybeta（见前文），来衡量在占中事件中港股对抗议活动的正向敏感度。并试图发现港股公司是否会从上次"教训"中吸取经验。

表 11 反应出 occupybeta 对累计异常收益率的影响。（1）至（2）列是仅针对经历过占中事件且在 2014 年前上市的 1479 只股票（为了计算 occupybeta）。由于其余的股票在 2014 年后上市，或没有经历过占中事件，（3）至（4）列的将 occupybeta 处理后生成变量 occupybeta'，把没有计算出 occupybeta 的股票的 occupybeta'值设为 0；可以计算出 occupybeta 的股票，occupybeta'值等于 occupybeta。（2）和（4）列结果表明，occupybeta（或 occupybeta'）与抗议人数的交叉项系数显著为负，即股票在占中事件中受到的抗议活动的负面冲击越大（occupybeta 值越小），在修例风波中受到抗议活动的冲击反而相对更小（AR 越大）。即证明了港股公司会从占中事件的抗议风波中吸收经验，受到冲击越大的公司在经过调整后，在修例风波中受到抗议带来的冲击越小。

为了进一步控制公司本身对抗议活动的敏感性，本文将处理后的 occupybeta'作为控制变量，进行面板回归，检验结论的稳健性。结果发现（表 12），结论与前文（表 8）相同：即在全时间段内，受到抗议活动额外冲击的主要为与泛民派有关的股票和中资民营股。而抗议活动对红筹股、H 股和内地政府控股的公司较其他股票没有显著的冲击。

红筹股在修例风波全时间段内有比其他股票更好的异常收益率。

## （二）其他考量

由于本事件属于政治产生的外生冲击，因此反向因果的关系较小，一部分内生性可以得到避免。然而，在事件时间段内，也有其他的外生因素会对香港股市产生影响。其中，中美贸易战会对股市产生较大的冲击。为了避免中美贸易战的影响，本文使用了异常收益率来消除系统性风险，并使用了 worldchange 变量（每日 MSCI 世界指数涨跌幅）来衡量世界经济、股市的走向。

为了进一步控制中美贸易战的外生影响，表 13 使用了 shindex 变量替代 worldchange（每日上证指数涨跌幅）作为控制变量。结果与表 8 相同（使用 worldchange 衡量世界经济的影响）。说明本文结论不会受到中美贸易战的影响而改变。

# 五、 结论

本文使用了事件研究法和面板回归模型，试图解答一个问题：有不同政治联系的香港上市公司会如何受到政治风险的影响？

通过事件研究法，本文发现，当有止暴制乱的政策消息出现时，与泛民派相关的股票会受到负面冲击，与内地有关联的股票会受到正面反馈。而出现政府对抗议抗议游行让步的信息时，中资民营股会受到负面冲击，与泛民派相关的股价估值将提高。

本文发现，在与香港政党有政治联系的公司中，与泛民派关联的公司受到了抗议活动的负面冲击。这种负面影响主要来源于香港特区政府明确表示要用法律手段止暴制乱之后，反映出了泛民派产生的政治联系具有负面价值。而与建制派相关的公司没有受到抗议活动的额外影响。

此外，本文发现，与内地相关联的公司受到了不同的影响。在《反蒙面法》实施之前，与内地相关联的公司普遍受到抗议活动的额外负面影响，本文认为是市场在修例风波初期产生巨大的不确定性，投资者对与内地相关联的公司产生了悲观情绪。在政府使用法律手段止暴制乱之后，市场的判断逻辑发生了改变。对未来政府进一步止暴制乱的期待体现在了对红筹股的额外关注。因此，红筹股受到了额外的正面反馈，体现出了红筹股政治联系的正面价值。

# 参考文献

报,1996(8).

# 附录

表 1　变量名称及定义

| 变量名 | 变量定义 | 单位 |
|---|---|---|
| **因变量** | | |
| AR | 异常收益率（计算方法见文中） | % |
| CAR | 累计异常收益率（计算方法见文中） | % |
| **自变量** | | |
| protests | 反对《逃犯条例》的抗议人数 | 人 |
| stdprotests | protests 变量经过标准化后 | 数值 |
| proestablish | 如果董监高成员中有至少一名是建制派政党党员，值取 1 否则取 0 | dummy |
| pandemo | 如果董监高成员中有至少一名是泛民派政党党员，值取 1 否则取 0 | dummy |
| H | H 股，如果公司在内地注册且在香港上市，值取 1 | dummy |
| red | 如果股票属于红筹股，值取 1（红筹股定义见文中） | dummy |
| centralcontrol | 如果公司被中央或地方政府，发改委控股，值取 1 | dummy |
| chinaasset | 如果股票属于中资民营股（非 H 股），值取 1（定义见文中） | dummy |
| occupybeta | 股票对抗议活动的敏感性（构建方法见文中） | $10^{-4}$ |
| **控制变量和其他变量** | | |
| size | 总资产对数 | 港元 |
| leverage | 资产负债率 | 数值 |
| 1/PE | 市盈率的倒数 | 数值 |
| Turnover | 换手率=每日交易量/总流通股数 | 数值 |
| hkbeta | 以 MSCI 香港指数为市场指数，股票的$\beta$ | 数值 |
| worldbeta | 以 MSCI 世界指数为市场指数，股票的$\beta$ | 数值 |
| hkchange | MSCI 香港指数每日涨跌幅 | % |
| worldchange | MSCI 世界指数每日涨跌幅 | % |
| shindex | 每日上证综指涨跌幅 | % |

表 2　数据统计性描述

| Panel A | (1) | (2) | (3) | (4) | (5) |
|---|---|---|---|---|---|
| Cross-section variables | N | mean | sd | min | max |
| | | | | | |
| proestablish | 1,961 | 0.158 | 0.364 | 0 | 1 |
| pandemo | 1,961 | 0.0367 | 0.188 | 0 | 1 |
| H | 1,961 | 0.123 | 0.328 | 0 | 1 |
| red | 1,961 | 0.0806 | 0.272 | 0 | 1 |
| centralcontrol | 1,961 | 0.141 | 0.348 | 0 | 1 |
| chinaasset | 2,047 | 0.275 | 0.447 | 0 | 1 |
| worldbeta | 1,961 | 0.411 | 0.420 | -2.920 | 2.222 |
| occupybeta | 1,479 | -0.0549 | 0.158 | -1.363 | 2.287 |
| size | 1,961 | 21.90 | 2.359 | 15.27 | 30.95 |
| leverage | 1,961 | 0.442 | 0.240 | 0.0004 | 1.000 |
| Number of code | 1,961 | 1,961 | 1,961 | 1,961 | 1,961 |

| Panel B | (1) | (2) | (3) | (4) | (5) |
|---|---|---|---|---|---|
| Timeseries variables | N | mean | sd | min | max |
| | | | | | |
| protests | 155 | 51,885 | 157,177 | 0 | 1170020 |
| stdprotests | 155 | 0 | 1.003 | -0.331 | 7.137 |
| hkchange | 155 | 0.0312 | 1.159 | -3.316 | 5.446 |
| worldchange | 157 | 0.0919 | 0.586 | -2.520 | 1.514 |
| shindex | 155 | 0.031 | 0.799 | -2.58 | 2.58 |

| Panel C | (1) | (2) | (3) | (4) | (5) |
|---|---|---|---|---|---|
| Panel variables | N | mean | sd | min | max |
| | | | | | |
| AR | 298,633 | -0.0596 | 3.716 | -68.73 | 432.0 |
| 1/PE | 298,632 | -0.149 | 1.721 | -96.01 | 12.71 |
| turnover | 298,633 | 0.0848 | 0.148 | 0 | 1 |

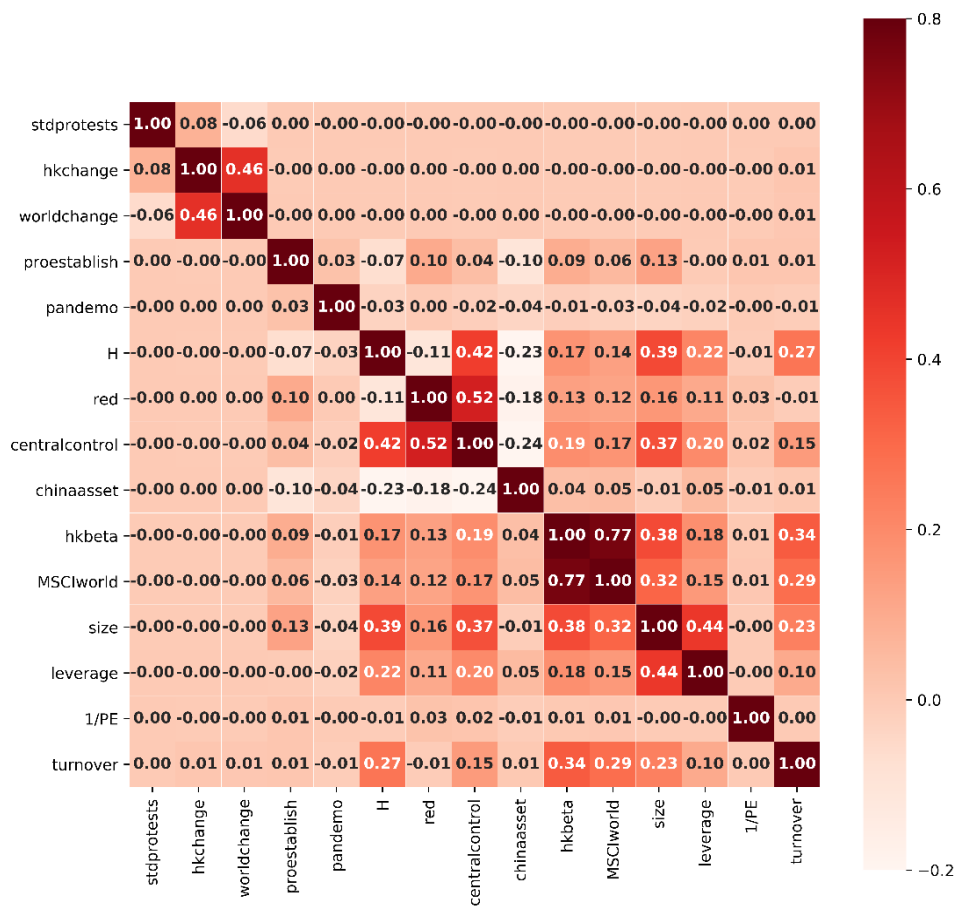

图 2　变量相关性

表 3　　修例风波中的重要政治事件

| 时间 | 事件 |
|---|---|
| 2019/6/15 | 特区政府决定暂缓修订《逃犯条例》的工作 |
| 2019/8/7 | 国务院港澳办和中央政府驻港联络办在深圳共同举办香港局势座谈会 |
| 2019/9/4 | 香港政府动议撤回《逃犯条例》修订 |
| 2019/10/5 | 开始实施《反蒙面法》 |
| 2019/10/23 | 保安局局长李家超宣布正式撤回《逃犯条例》修订草案 |
| 2019/11/4 | 国家主席习近平在上海会见香港特别行政区行政长官林郑月娥 |
| 2019/11/18 | 香港特别行政区高等法院裁定香港《紧急情况规例条例》、《反蒙面法》无效 |
| 2019/12/6 | 公安部部长赵克志在京会见香港警务处处长邓炳强 |

表 4 《反蒙面法》实施的事件分析结果

| VARIABLES | (1) | (2) | (3) | (4) | (5) | (6) |
|---|---|---|---|---|---|---|
| | CAR[83,87] | | | CAR[84,86] | | |
| Proestablish | -0.037 | | | -0.200 | | |
| | (0.925) | | | (0.554) | | |
| Pandemo | -1.974*** | | | -1.123* | | |
| | (0.009) | | | (0.083) | | |
| H | | 0.787 | | | 0.902** | |
| | | (0.112) | | | (0.032) | |
| red | | 1.324** | | | 1.131** | |
| | | (0.015) | | | (0.015) | |
| centralcontrol | | | 0.928** | | | 0.764* |
| | | | (0.048) | | | (0.055) |
| chinaasset | | | 0.736** | | | 0.459 |
| | | | (0.029) | | | (0.110) |
| MSCIworld | 2.257*** | 2.205*** | 2.199*** | 1.812*** | 1.757*** | 1.762*** |
| | (0.000) | (0.000) | (0.000) | (0.000) | (0.000) | (0.000) |
| size | 0.322*** | 0.269*** | 0.282*** | 0.265*** | 0.202*** | 0.226*** |
| | (0.000) | (0.001) | (0.000) | (0.000) | (0.002) | (0.001) |
| leverage | -0.192 | -0.301 | -0.320 | -0.421 | -0.505 | -0.482 |
| | (0.775) | (0.654) | (0.634) | (0.461) | (0.376) | (0.399) |
| Constant | -9.014*** | -8.135*** | -8.427*** | -6.914*** | -5.800*** | -6.311*** |
| | (0.000) | (0.000) | (0.000) | (0.000) | (0.003) | (0.001) |
| Observations | 1,944 | 1,944 | 1,944 | 1,943 | 1,943 | 1,943 |
| R-squared | 0.058 | 0.059 | 0.058 | 0.049 | 0.051 | 0.049 |
| Control Variables | YES | YES | YES | YES | YES | YES |
| industry effect | YES | YES | YES | YES | YES | YES |

表 5　《逃犯条例》正式撤回的事件分析结果

| VARIABLES | (1) | (2) | (3) | (4) | (5) | (6) |
|---|---|---|---|---|---|---|
| | CAR[94,98] | | | CAR[95,97] | | |
| proestablish | 0.162 | | | -0.101 | | |
| | (0.698) | | | (0.753) | | |
| pandemo | 1.548* | | | 1.812*** | | |
| | (0.053) | | | (0.003) | | |
| H | | -0.039 | | | 0.115 | |
| | | (0.940) | | | (0.775) | |
| red | | 0.526 | | | 0.018 | |
| | | (0.359) | | | (0.967) | |
| centralcontrol | | | -0.245 | | | -0.307 |
| | | | (0.619) | | | (0.418) |
| chinaasset | | | -0.995*** | | | -0.733*** |
| | | | (0.005) | | | (0.007) |
| MSCIworld | -0.641* | -0.687* | -0.597 | -0.326 | -0.350 | -0.299 |
| | (0.095) | (0.074) | (0.120) | (0.269) | (0.237) | (0.312) |
| size | 0.033 | 0.030 | 0.043 | 0.012 | 0.001 | 0.020 |
| | (0.670) | (0.713) | (0.597) | (0.835) | (0.983) | (0.752) |
| leverage | -0.404 | -0.445 | -0.290 | 0.231 | 0.228 | 0.343 |
| | (0.567) | (0.529) | (0.681) | (0.670) | (0.674) | (0.527) |
| Constant | -0.670 | -0.513 | -0.531 | -0.211 | 0.025 | -0.248 |
| | (0.750) | (0.815) | (0.805) | (0.907) | (0.989) | (0.892) |
| Observations | 1,941 | 1,941 | 1,941 | 1,939 | 1,939 | 1,939 |
| R-squared | 0.008 | 0.007 | 0.010 | 0.011 | 0.007 | 0.010 |
| Control Variables | YES | YES | YES | YES | YES | YES |
| industry effect | YES | YES | YES | YES | YES | YES |

pval in parentheses

*** p<0.01, ** p<0.05, * p<0.1

表6  公安部部长会见香港警务处处长事件研究结果

| VARIABLES | (1) | (2) | (3) | (4) | (5) | (6) |
|---|---|---|---|---|---|---|
| | CAR[126,130] | | | CAR[127,129] | | |
| proestablish | 0.297 | | | 0.199 | | |
| | (0.472) | | | (0.565) | | |
| pandemo | 0.602 | | | -0.513 | | |
| | (0.443) | | | (0.435) | | |
| H | | 0.962* | | | 0.138 | |
| | | (0.061) | | | (0.749) | |
| red | | 1.223** | | | 0.665 | |
| | | (0.031) | | | (0.161) | |
| centralcontrol | | | 0.841* | | | -0.067 |
| | | | (0.084) | | | (0.869) |
| chinaasset | | | 0.056 | | | -0.302 |
| | | | (0.874) | | | (0.305) |
| MSCIworld | 0.616 | 0.542 | 0.570 | -0.268 | -0.296 | -0.240 |
| | (0.103) | (0.152) | (0.133) | (0.399) | (0.353) | (0.451) |
| size | 0.111 | 0.052 | 0.075 | 0.043 | 0.034 | 0.051 |
| | (0.149) | (0.519) | (0.345) | (0.501) | (0.623) | (0.449) |
| leverage | -0.226 | -0.374 | -0.301 | -0.112 | -0.168 | -0.094 |
| | (0.745) | (0.591) | (0.665) | (0.848) | (0.773) | (0.872) |
| Constant | -2.600 | -1.294 | -1.807 | 1.335 | 1.423 | 1.229 |
| | (0.210) | (0.549) | (0.394) | (0.490) | (0.475) | (0.530) |
| | | | | | | |
| Observations | 1,937 | 1,937 | 1,937 | 1,933 | 1,933 | 1,933 |
| R-squared | 0.011 | 0.014 | 0.012 | 0.013 | 0.014 | 0.013 |
| Control Variables | YES | YES | YES | YES | YES | YES |
| industry effect | YES | YES | YES | YES | YES | YES |

pval in parentheses

*** p<0.01, ** p<0.05, * p<0.1

表 7　修例风波中不同种类股票的累计异常收益率

| VARIABLES | (1) | (2) | (3) | (4) | (5) | (6) | (7) | (8) | (9) |
|---|---|---|---|---|---|---|---|---|---|
| | CAR[0,155] | | | CAR[0,84] | | | CAR[85,155] | | |
| proestablish | 4.368* | | | 1.848 | | | 2.521* | | |
| | (0.056) | | | (0.272) | | | (0.080) | | |
| pandemo | -5.315 | | | -3.836 | | | -1.476 | | |
| | (0.227) | | | (0.236) | | | (0.594) | | |
| H | | 4.138 | | | 2.841 | | | 1.298 | |
| | | (0.125) | | | (0.153) | | | (0.445) | |
| red | | 7.905** | | | 4.057* | | | 3.848* | |
| | | (0.012) | | | (0.079) | | | (0.051) | |
| centralcontrol | | | 5.524** | | | 3.365* | | | 2.159 |
| | | | (0.034) | | | (0.079) | | | (0.188) |
| chinaasset | | | -1.513 | | | -0.078 | | | -1.433 |
| | | | (0.439) | | | (0.957) | | | (0.245) |
| Constant | -20.402** | -21.479** | -21.110** | -5.069 | -5.612 | -5.624 | -15.339*** | -15.871*** | -15.491*** |
| | (0.028) | (0.021) | (0.023) | (0.458) | (0.412) | (0.411) | (0.009) | (0.007) | (0.008) |
| Observations | 1,948 | 1,948 | 1,948 | 1,948 | 1,948 | 1,948 | 1,947 | 1,947 | 1,947 |
| R-squared | 0.009 | 0.010 | 0.009 | 0.018 | 0.019 | 0.019 | 0.013 | 0.013 | 0.013 |
| Control Variables | YES | YES | YES | YES | YES | YES | YES | YES | YES |
| industry effect | YES | YES | YES | YES | YES | YES | YES | YES | YES |

表 8　抗议活动剧烈程度对各类有政治联系的香港股票的影响

| VARIABLES | (1) | (2) | (3) | (4) AR | (5) | (6) | (7) |
|---|---|---|---|---|---|---|---|
| stdprotests | -0.035*** | -0.035*** | -0.029*** | -0.035*** | -0.032*** | -0.035*** | -0.019** |
| | (0.000) | (0.000) | (0.000) | (0.000) | (0.000) | (0.000) | (0.041) |
| proestablish | | 0.021 | 0.021 | | | | |
| | | (0.261) | (0.273) | | | | |
| pandemo | | -0.037 | -0.038 | | | | |
| | | (0.316) | (0.295) | | | | |
| stdprotests*proestablish | | | -0.021 | | | | |
| | | | (0.281) | | | | |
| stdprotests* pandemo | | | -0.065* | | | | |
| | | | (0.085) | | | | |
| H | | | | -0.126*** | -0.126*** | | |
| | | | | (0.000) | (0.000) | | |
| red | | | | 0.045* | 0.044* | | |
| | | | | (0.082) | (0.089) | | |
| stdprotests* H | | | | | 0.003 | | |
| | | | | | (0.881) | | |
| stdprotests* red | | | | | -0.040 | | |
| | | | | | (0.123) | | |
| centralcontrol | | | | | | -0.029 | -0.030 |
| | | | | | | (0.195) | (0.185) |
| chinaasset | | | | | | -0.018 | -0.019 |
| | | | | | | (0.262) | (0.233) |
| stdprotests*central | | | | | | | -0.026 |
| | | | | | | | (0.221) |
| stdprotests*chinaasset | | | | | | | -0.044*** |
| | | | | | | | (0.007) |
| Constant | 0.089 | 0.097 | 0.098 | -0.014 | -0.014 | 0.064 | 0.064 |
| | (0.210) | (0.171) | (0.170) | (0.854) | (0.854) | (0.391) | (0.388) |
| Observations | 286,133 | 286,133 | 286,133 | 286,133 | 286,133 | 286,133 | 286,133 |
| Number of code | 1,961 | 1,961 | 1,961 | 1,961 | 1,961 | 1,961 | 1,961 |
| Control variables | YES | YES | YES | YES | YES | YES | YES |
| Industry effect | YES | YES | YES | YES | YES | YES | YES |

pval in parentheses
*** p<0.01, ** p<0.05, * p<0.1

表 9 《反蒙面法》实施前抗议活动对各类有政治联系的股票的影响

| VARIABLES | (1) | (2) | (3) | (4) | (5) | (6) | (7) |
|---|---|---|---|---|---|---|---|
| | | | | 《反蒙面法》实施之前 | | | |
| | | | | AR | | | |
| stdprotests | -0.050*** | -0.050*** | -0.045*** | -0.050*** | -0.040*** | -0.050*** | -0.028*** |
| | (0.000) | (0.000) | (0.000) | (0.000) | (0.000) | (0.000) | (0.007) |
| proestablish | | 0.021 | 0.022 | | | | |
| | | (0.426) | (0.393) | | | | |
| pandemo | | -0.055 | -0.051 | | | | |
| | | (0.275) | (0.307) | | | | |
| stdprotests*proestablish | | | -0.020 | | | | |
| | | | (0.354) | | | | |
| stdprotests* pandemo | | | -0.043 | | | | |
| | | | (0.315) | | | | |
| H | | | | -0.115*** | -0.114*** | | |
| | | | | (0.001) | (0.001) | | |
| red | | | | 0.052 | 0.059* | | |
| | | | | (0.144) | (0.096) | | |
| stdprotests* H | | | | | -0.021 | | |
| | | | | | (0.406) | | |
| stdprotests* red | | | | | -0.096*** | | |
| | | | | | (0.001) | | |
| centralcontrol | | | | | | -0.018 | -0.014 |
| | | | | | | (0.558) | (0.656) |
| chinaasset | | | | | | -0.007 | -0.004 |
| | | | | | | (0.736) | (0.872) |
| stdprotests*central | | | | | | | -0.057** |
| | | | | | | | (0.016) |
| stdprotests*chinaasset | | | | | | | -0.050*** |
| | | | | | | | (0.006) |
| Constant | 0.055 | 0.065 | 0.065 | -0.036 | -0.037 | 0.039 | 0.037 |
| | (0.570) | (0.505) | (0.506) | (0.729) | (0.723) | (0.703) | (0.716) |
| | | | | | | | |
| Observations | 152,844 | 152,844 | 152,844 | 152,844 | 152,844 | 152,844 | 152,844 |
| Number of code | 1,961 | 1,961 | 1,961 | 1,961 | 1,961 | 1,961 | 1,961 |
| Control variables | YES | YES | YES | YES | YES | YES | YES |
| Industry effect | YES | YES | YES | YES | YES | YES | YES |

pval in parentheses

*** p<0.01, ** p<0.05, * p<0.1

表 10 《反蒙面法》实施后抗议活动对各类有政治联系的股票的影响

| VARIABLES | (1) | (2) | (3) | (4) | (5) | (6) | (7) |
|---|---|---|---|---|---|---|---|
| | | | | 《反蒙面法》实施之后 AR | | | |
| stdprotests | 0.039** | 0.039** | 0.047*** | 0.039** | 0.017 | 0.039** | 0.039* |
| | (0.016) | (0.016) | (0.009) | (0.016) | (0.351) | (0.016) | (0.062) |
| proestablish | | 0.025 | 0.022 | | | | |
| | | (0.369) | (0.445) | | | | |
| pandemo | | 0.000 | -0.018 | | | | |
| | | (0.999) | (0.740) | | | | |
| stdprotests*proestablish | | | -0.022 | | | | |
| | | | (0.622) | | | | |
| stdprotests* pandemo | | | -0.123* | | | | |
| | | | (0.054) | | | | |
| H | | | | -0.147*** | -0.135*** | | |
| | | | | (0.000) | (0.000) | | |
| red | | | | 0.031 | 0.056 | | |
| | | | | (0.421) | (0.161) | | |
| stdprotests* H | | | | | 0.076 | | |
| | | | | | (0.131) | | |
| stdprotests* red | | | | | 0.163*** | | |
| | | | | | (0.006) | | |
| centralcontrol | | | | | | -0.050 | -0.039 |
| | | | | | | (0.139) | (0.254) |
| chinaasset | | | | | | -0.034 | -0.039 |
| | | | | | | (0.157) | (0.110) |
| stdprotests*central | | | | | | | 0.069 |
| | | | | | | | (0.148) |
| stdprotests*chinaasset | | | | | | | -0.036 |
| | | | | | | | (0.327) |
| Constant | 0.204* | 0.210** | 0.211** | 0.079 | 0.075 | 0.162 | 0.162 |
| | (0.051) | (0.045) | (0.044) | (0.480) | (0.499) | (0.139) | (0.139) |
| Observations | 131,366 | 131,366 | 131,366 | 131,366 | 131,366 | 131,366 | 131,366 |
| Number of code | 1,961 | 1,961 | 1,961 | 1,961 | 1,961 | 1,961 | 1,961 |
| Control variables | YES | YES | YES | YES | YES | YES | YES |
| Industry effect | YES | YES | YES | YES | YES | YES | YES |

pval in parentheses

*** p<0.01, ** p<0.05, * p<0.1

表 11　港股对抗议活动的敏感性

| VARIABLES | (1) | (2) | (3) | (4) |
|---|---|---|---|---|
| | | | AR | |
| stdprotests | -0.025*** | -0.030*** | -0.035*** | -0.039*** |
| | (0.002) | (0.000) | (0.000) | (0.000) |
| occupybeta | -0.001 | -0.003 | | |
| | (0.984) | (0.951) | | |
| occupybeta* stdprotests | | -0.085* | | |
| | | (0.094) | | |
| occupybeta' | | | -0.015 | -0.017 |
| | | | (0.767) | (0.729) |
| occupybeta'* stdprotests | | | | -0.101** |
| | | | | (0.048) |
| worldchange | -0.156*** | -0.156*** | -0.161*** | -0.161*** |
| | (0.000) | (0.000) | (0.000) | (0.000) |
| size | -0.012*** | -0.012*** | -0.011*** | -0.011*** |
| | (0.004) | (0.004) | (0.003) | (0.003) |
| leverage | -0.045 | -0.045 | 0.001 | 0.001 |
| | (0.205) | (0.205) | (0.983) | (0.984) |
| 1/PE | 0.022*** | 0.022*** | 0.023*** | 0.023*** |
| | (0.000) | (0.000) | (0.000) | (0.000) |
| turnover | 1.308*** | 1.308*** | 1.323*** | 1.323*** |
| | (0.000) | (0.000) | (0.000) | (0.000) |
| Constant | 0.138 | 0.138 | 0.087 | 0.087 |
| | (0.118) | (0.119) | (0.221) | (0.222) |
| | | | | |
| Observations | 216,348 | 216,348 | 286,133 | 286,133 |
| Number of code | 1,479 | 1,479 | 1,961 | 1,961 |
| Control variables | YES | YES | YES | YES |
| Industry effect | YES | YES | YES | YES |

pval in parentheses

*** p<0.01, ** p<0.05, * p<0.1

表 12　加入 occupybeta'后的稳健性检验

| VARIABLES | (1) | (2) | (3) | (4) | (5) | (6) | (7) |
|---|---|---|---|---|---|---|---|
| | | | | AR | | | |
| stdprotests | -0.035*** | -0.035*** | -0.029*** | -0.035*** | -0.032*** | -0.035*** | -0.019** |
| | (0.000) | (0.000) | (0.000) | (0.000) | (0.000) | (0.000) | (0.041) |
| proestablish | | 0.021 | 0.021 | | | | |
| | | (0.262) | (0.274) | | | | |
| pandemo | | -0.037 | -0.038 | | | | |
| | | (0.316) | (0.295) | | | | |
| stdprotests*proestablish | | | -0.021 | | | | |
| | | | (0.281) | | | | |
| stdprotests* pandemo | | | -0.065* | | | | |
| | | | (0.085) | | | | |
| H | | | | -0.126*** | -0.126*** | | |
| | | | | (0.000) | (0.000) | | |
| red | | | | 0.045* | 0.044* | | |
| | | | | (0.082) | (0.090) | | |
| stdprotests* H | | | | | 0.003 | | |
| | | | | | (0.881) | | |
| stdprotests* red | | | | | -0.040 | | |
| | | | | | (0.123) | | |
| centralcontrol | | | | | | -0.029 | -0.030 |
| | | | | | | (0.198) | (0.188) |
| chinaasset | | | | | | -0.018 | -0.019 |
| | | | | | | (0.265) | (0.237) |
| stdprotests*central | | | | | | | -0.026 |
| | | | | | | | (0.221) |
| stdprotests*chinaasset | | | | | | | -0.044*** |
| | | | | | | | (0.007) |
| occupybeta' | -0.015 | -0.014 | -0.014 | 0.002 | 0.002 | -0.012 | -0.012 |
| | (0.767) | (0.773) | (0.773) | (0.962) | (0.961) | (0.810) | (0.810) |
| Constant | 0.087 | 0.095 | 0.096 | -0.014 | -0.014 | 0.062 | 0.063 |
| | (0.221) | (0.180) | (0.179) | (0.856) | (0.856) | (0.402) | (0.399) |
| Observations | 286,133 | 286,133 | 286,133 | 286,133 | 286,133 | 286,133 | 286,133 |
| Number of code | 1,961 | 1,961 | 1,961 | 1,961 | 1,961 | 1,961 | 1,961 |
| Control variables | YES | YES | YES | YES | YES | YES | YES |
| Industry effect | YES | YES | YES | YES | YES | YES | YES |

pval in parentheses

*** p<0.01, ** p<0.05, * p<0.1

表 13　控制中美贸易战的影响　（加入控制变量 shindex）

| VARIABLES | (1) | (2) | (3) | (4) | (5) | (6) | (7) |
|---|---|---|---|---|---|---|---|
| | | | | AR | | | |
| stdprotests | -0.018** | -0.018** | -0.012 | -0.018** | -0.016** | -0.018** | -0.004 |
| | (0.012) | (0.012) | (0.132) | (0.012) | (0.040) | (0.012) | (0.693) |
| shindex | -0.031*** | -0.031*** | -0.031*** | -0.031*** | -0.031*** | -0.031*** | -0.031*** |
| | (0.000) | (0.000) | (0.000) | (0.000) | (0.000) | (0.000) | (0.000) |
| proestablish | | 0.018 | 0.017 | | | | |
| | | (0.352) | (0.363) | | | | |
| pandemo | | -0.035 | -0.036 | | | | |
| | | (0.343) | (0.327) | | | | |
| stdprotests*proestablish | | | -0.023 | | | | |
| | | | (0.239) | | | | |
| stdprotests* pandemo | | | -0.068* | | | | |
| | | | (0.070) | | | | |
| H | | | | -0.112*** | -0.112*** | | |
| | | | | (0.000) | (0.000) | | |
| red | | | | 0.049* | 0.049* | | |
| | | | | (0.061) | (0.064) | | |
| stdprotests* H | | | | | 0.010 | | |
| | | | | | (0.639) | | |
| stdprotests* red | | | | | -0.035 | | |
| | | | | | (0.177) | | |
| centralcontrol | | | | | | -0.023 | -0.023 |
| | | | | | | (0.319) | (0.312) |
| chinaasset | | | | | | -0.021 | -0.022 |
| | | | | | | (0.204) | (0.188) |
| stdprotests*central | | | | | | | -0.020 |
| | | | | | | | (0.353) |
| stdprotests*chinaasset | | | | | | | -0.042** |
| | | | | | | | (0.011) |
| Constant | 0.086 | 0.093 | 0.094 | -0.003 | -0.003 | 0.068 | 0.068 |
| | (0.228) | (0.191) | (0.190) | (0.973) | (0.973) | (0.362) | (0.360) |

| Observations | 284,174 | 284,174 | 284,174 | 284,174 | 284,174 | 284,174 | 284,174 |
|---|---|---|---|---|---|---|---|
| Number of code | 1,961 | 1,961 | 1,961 | 1,961 | 1,961 | 1,961 | 1,961 |
| Control variables | YES | YES | YES | YES | YES | YES | YES |
| Industry effect | YES | YES | YES | YES | YES | YES | YES |

pval in parentheses

*** p<0.01, ** p<0.05, * p<0.1